\documentclass[twocolumn,letter]{jpsj3}
\usepackage{bm}
\usepackage{graphicx}
\usepackage{subfigure}
\usepackage[usenames]{color}

\title{Single Impurity Problem in Iron-Pnictide Superconductors}
\author{Toshikaze \textsc{Kariyado}$^1$\thanks{E-mail: kariyado@hosi.phys.s.u-tokyo.ac.jp} 
and Masao \textsc{Ogata}$^{1,2}$}
\inst{$^1$Department of Physics, University of Tokyo, 7-3-1 Hongo, Bunkyo, Tokyo
113-0033\\
$^2$JST, TRIP, Sanbancho, Chiyoda, Tokyo 102-0075}

\abst{
Single impurity problem in iron-pnictide superconductors is
investigated by solving Bogoliubov-de Gennes (BdG) equation in the
five-orbital model, which enables us to distinguish s$_{+-}$ and
s$_{++}$ superconducting states. We construct a five-orbital model
suitable to BdG analysis. This model reproduces the results of random
phase approximation in the uniform case. Using this model, we study the
local density of states
around a non-magnetic impurity and discuss the bound-state peak
structure, which can be used for distinguishing s$_{+-}$ and s$_{++}$
states. A bound state with nearly zero-energy is found for the impurity
potential $I\sim 1.0$ eV, while the bound state peaks stick to the gap
edge in the unitary limit. Novel multiple peak structure originated from
the multi-orbital nature of the iron pnictides is also found.
}

\kword{iron pnictides, superconductivity, Bogoliubov-de Gennes equation,
impurity}

\usepackage{amsmath}	
\begin{document}
\maketitle

Much effort has been devoted to elucidate the structure of
superconducting gap
functions in the recently discovered iron
pnictides\cite{JPSJ.78.062001}. However, there are
still many debates on this issue. Experimentally, the measurement
of the quasi-particle interference pattern\cite{hanaguri} and the
existence of the half-integer flux-quantum transition\cite{Chen:2009e} suggest
that there is an internal $\pi$-phase shift of the gap function in
iron-pnictide superconductors. Furthermore, nodal behaviors observed
in some families of iron pnictides\cite{Hashimoto:2009} indicate that
the repulsive interaction plays important roles and
the pairing with $\pi$-phase shift of the gap function sounds
reasonable. On the other hand, it is claimed that the robustness against
impurity doping cannot be consistent with 
the above mechanism\cite{Sato:2009,Onari:2009}.
Many theoretical studies have been performed in order to explain the
high temperature
superconductivity in iron
pnictides\cite{Mazin:2008,Kuroki:2008,Chubukov:2008,Yanagi:2008b,Ikeda:2008,Fuseya:2009,Cvetkovic:2008,Graser:2009,Nomura:2008b,JPSJ.79.033703,Kontani:2009}. Most
of these theories have concluded
that there exists inter-band sign reversal of the gap function (s$_{+-}$
state). Recently, high temperature superconductivity without
inter-band sign reversal (s$_{++}$ state) is also suggested\cite{Kontani:2009}. 
Considering these situations, it is important and urgent to develop some
theories which enable us to distinguish the s$_{+-}$ and s$_{++}$
states. There have been several theoretical proposals for detecting the
s$_{+-}$ state. However, it is a difficult and subtle problem to
distinguish s$_{+-}$ and s$_{++}$ state than to detect other
unconventional superconducting states such as d-wave pairing, since the
symmetries of the gap functions are the same for s$_{+-}$ and s$_{++}$ state. 

In this paper, we investigate a single impurity
problem since it can serve as a possible method to detect the sign
change of the gap function. Actually, this problem has been discussed
mainly in some simplified two band models for iron-pnictide
superconductors\cite{PhysRevB.79.054529,Matsumoto:2009,Tsai:2009,Ng:2009,Li:2009,Zhang:2009,Akbari:2010,Zhou:2009}. However,
as we show here, 
detecting the s$_{+-}$ state is subtle problem and the entangled nature
of the multiple bands and the Fermi surface
structure, which characterize iron pnictides, should be taken into
account. In the first part of this paper, we construct a model which
can be used in the real-space single impurity problem. Although the
five-orbital {\it Hubbard} model has been studied quite often, it is not
suitable for our purpose since it does not give a superconducting ground
state in the mean field theory. Here we construct a model whose
interaction terms are chosen so as to 
reproduce the results in the five-orbital Hubbard model calculated within
random phase approximation (RPA). The hopping integrals of this model
are taken from the downfolded result. Then, the local density of states
(LDOS) around a single non-magnetic impurity is calculated by solving
Bogoliubov-de Gennes (BdG) equation for the obtained model. The results show
clear formation of in-gap bound state, which turns out to be a good
quantity for distinguishing s$_{+-}$ and s$_{++}$ states. We find that
the dependence
of the spectrum on the impurity potential strength is unique to the
present system, and there are novel multiple peak structures in LDOS for
a certain parameter range.

Now, we explain the five-orbital model used in this study. Hamiltonian
for the used model can be written as
\begin{eqnarray}
 \mathcal{H}&=&\sum_{\bm{i}\bm{j}}\sum_{\sigma}\sum_{ab}
  t_{a,b;\bm{i},\bm{j}}c^\dagger_{\bm{i}a\sigma}c_{\bm{j}b\sigma}
  +I\sum_{\sigma}\sum_{a}c^\dagger_{\bm{r}^*a\sigma}
  c_{\bm{r}^*a\sigma}\nonumber\\
  &+&\sum_{\bm{i}}\sum_{\sigma}\sum_{aba'b'}g^{(0)}_{aa'b'b}
  c^\dagger_{\bm{i}a\sigma}c^\dagger_{\bm{i}b\bar{\sigma}}
  c_{\bm{i}a'\bar{\sigma}}c_{\bm{i}b'\sigma}\nonumber\\
  &+&\sum_{\langle\bm{i}\bm{j}\rangle}\sum_{\sigma}\sum_{aba'b'}
   g^{(1)}_{aa'b'b}
  c^\dagger_{\bm{i}a\sigma}c^\dagger_{\bm{j}b\bar{\sigma}}
  c_{\bm{j}a'\bar{\sigma}}c_{\bm{i}b'\sigma}\nonumber\\
 &+&\sum_{\langle\langle\bm{i}\bm{j}\rangle\rangle}
  \sum_{\sigma}\sum_{aba'b'}g^{(2)}_{aa'b'b}
  c^\dagger_{\bm{i}a\sigma}c^\dagger_{\bm{j}b\bar{\sigma}}
  c_{\bm{j}a'\bar{\sigma}}c_{\bm{i}b'\sigma},
  \label{Hamiltonian}
\end{eqnarray}
where $r^*$ represents the position of the impurity site, and 
$\langle \bm{i}\bm{j}\rangle$ and 
$\langle\langle \bm{i}\bm{j}\rangle\rangle$ represent the
nearest and next nearest neighbor pairs respectively. Indices $a$ or $b$ run
through 0 to 4, where 0 to 4 correspond to $d_{3z^2-r^2}$, $d_{zx}$, $d_{yz}$,
$d_{x^2-y^2}$, and $d_{xy}$ orbitals in this order. Hopping integrals
$t_{a,b;\bm{i},\bm{j}}$ are same as those in the Table~I of Kuroki 
{\it et al}\cite{Kuroki:2008}. For details of these hopping parameters,
see Ref 8) and a brief description is also
available in our previours paper\cite{JPSJ.79.033703}. The effect of
impurity is simply treated as a
local potential shift, $I$, in eq.~(\ref{Hamiltonian}).
The orbital dependence of the potential shift and the
long-range impurity effects are neglected here. 

Coupling constants $g^{(i)}$s in eq.~(\ref{Hamiltonian}) are chosen so
as to reproduce the superconducting gap functions in the multi-orbital
Hubbard model calculated within
RPA\cite{JPSJ.79.033703}. It was shown that the gap function
obtained in RPA can be well reproduced with short-range pairings up to the
next nearest neighbor sites when it is written in the {\it orbital}
representation, instead of the {\it band} representation. This point is
reflected in eq.~(\ref{Hamiltonian}), where it is written in the orbital
representation and the interaction terms are kept up to the next nearest
neighbor sites. It was also shown that the pairings of $d_{3z^2-r^2}$
and $d_{xy}$ orbitals are less important
since the density of states (DOS) just around the Fermi energy mainly
comes from $d_{zx/yz}$ and $d_{x^2-y^2}$
orbitals\cite{JPSJ.79.033703}. Furthermore, the inter-orbital pairings
are less important than the intra-orbital
ones in the s-wave channel. Based on these features, we have reduced the
number of necessary parameters. Using the effective interaction
calculated within RPA ($V^{\mathrm{RPA}}_{abcd}(k)$, not shown), we obtain
\begin{subequations}
 \begin{eqnarray}
 &&g^{(0)}_{1111}=g^{(0)}_{2222}=g^{(0)}_{3333}=3.0,\\
 &&g^{(1)}_{1111}=g^{(1)}_{2222}=g^{(1)}_{3333}=-0.75,\\
 &&g^{(2)}_{1111}=g^{(2)}_{2222}=g^{(2)}_{3333}=-0.27,\\
 &&g^{(0)}_{1221}=g^{(0)}_{1331}=g^{(0)}_{2332}=0.3,\\
 &&g^{(1)}_{1221}=g^{(1)}_{1331}=g^{(1)}_{2332}=0.075,\\
 &&g^{(2)}_{1221}=g^{(2)}_{1331}=g^{(2)}_{2332}=-0.027,
 \end{eqnarray}
\end{subequations}
in the unit of eV. Note that $g^{(i)}_{abba}=g^{(i)}_{baab}$ and
parameters not shown here are set to be zero. $g^{(0)}_{aaaa}$
represents the onsite repulsion and $g^{(1)}_{aaaa}$ ($g^{(2)}_{aaaa}$)
represents the effective attractive interaction between the nearest
(next nearest) neighbor sites, mainly induced by the broad $(\pi,\pi)$
(sharp near $(\pi,0)$) peak in $V^{\mathrm{RPA}}_{aaaa}(k)$. 

\begin{figure}[tb]
\begin{center}
 \subfigure[$\Delta^{\mathrm{n.n.n.}}_{11}(\bm{r})$]
 {\includegraphics[width=120pt]{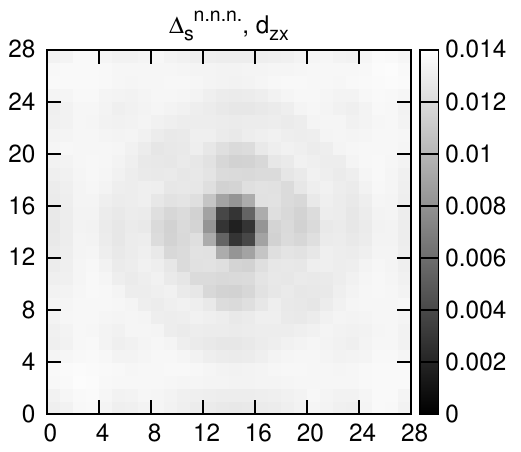}}
 \subfigure[$\Delta^{\mathrm{n.n.n.}}_{33}(\bm{r})$]
 {\includegraphics[width=120pt]{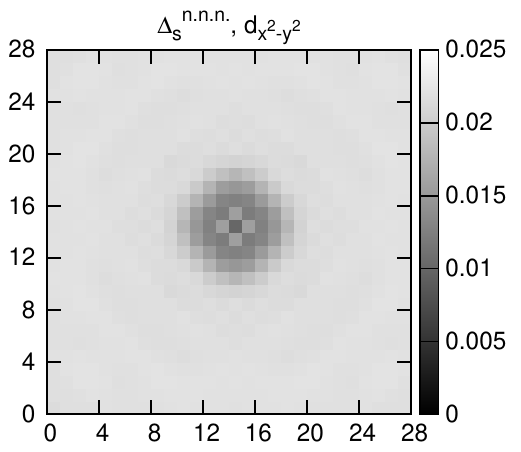}}
 \caption{Real space distribution of the order parameter 
$\Delta^{\mathrm{n.n.n.}}_{aa}(\bm{r})$ around the impurity
 site with $I=1.0$ eV. (a) for the orbital 1 ($d_{zx}$) and (b) for
 the orbital 3 ($d_{x^2-y^2}$)\cite{JPSJ.79.033703}.}
 \label{fig1}
\end{center}
\end{figure}

Hamiltonian, eq.~(\ref{Hamiltonian}), is solved in the mean field
approximation, where only the BCS type decomposition is used neglecting
the Hartree-Fock type contribution. Order parameters are assumed to be
site-dependent and then we obtain the multi-orbital version of the BdG
equation. In actual calculations, we introduce the energy cutoff,
$\omega_c$, necessary in the standard BCS theory, i.e., we assume that the
pairing occurs only between the electrons with energies $\varepsilon$
satisfying $|\varepsilon-\mu|<\omega_c$, where $\mu$ is the chemical
potential. Throughout this paper, we use $\omega_c=0.1$ eV.
This choice of $\omega_c$ does not affect the following discussion. We
solve the BdG equation in the lattices up to 28$\times$28 sites with an
impurity at the center, and determine the site-dependent order
parameters by iteration. The electron number is fixed at $n=6.1$ by
adjusting the chemical potantial, $\mu$. In order to calculate physical
quantities, such as LDOS, we use the ``super-cell''
method\cite{JPSJ.66.3367} to improve the numerical accuracy. In this
``super-cell'' method, the 28$\times$28 lattice with an impurity is
treated as a ``unit cell'' and the whole system is composed of
13$\times$13 repetition of this unit cell. The connection beween the
unit cells is represented by the wave vectors $\bm{k}=(2\pi n_x/13,2\pi
n_y/13)$. The LDOS in this ``super-cell'' method is given by
\begin{eqnarray}
 \rho(\bm{r},\omega)&=&\sum_a\rho_a(\bm{r},\omega),\\
  \rho_a(\bm{r},\omega)&=&
   -\frac{1}{\pi}\sum_{\bm{k}}\mathrm{Im}G^{R}_{\bm{r}a,\bm{r}a}(\bm{k},\omega)
\end{eqnarray}
where $\rho_a(\bm{r},\omega)$ is the partial LDOS for the orbital $a$,
and $G^R_{\bm{r}a,\bm{r}'a'}(\bm{k},\omega)$ is the retarded Green's
function in real space for the lattice sites $\bm{r}$ and $\bm{r}'$ in
the unit cell. In following figures, the Dirac delta functions in the
LDOS are replaced by the lorentzian function with the half width
$\gamma=0.001$ eV.  

\begin{figure}[tb]
 \begin{center}
  \subfigure[onsite]{\includegraphics[width=120pt]{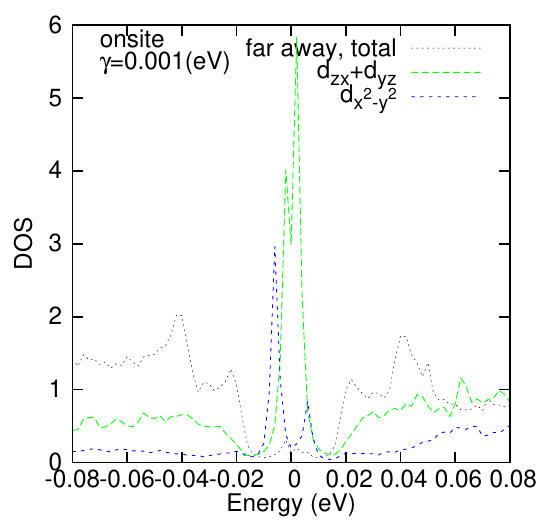}}
  \subfigure[n.n.]{\includegraphics[width=120pt]{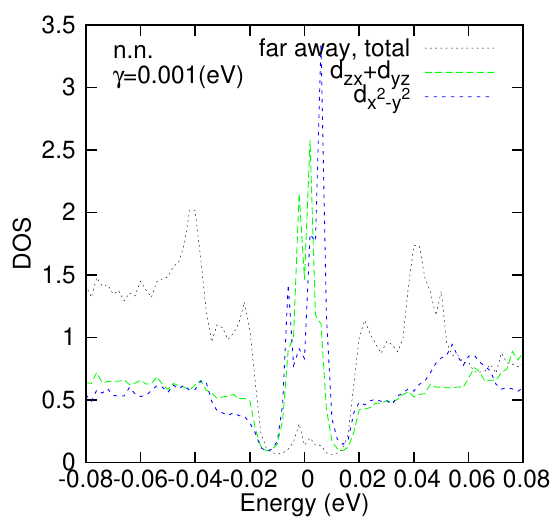}}
  \subfigure[n.n.n.]{\includegraphics[width=120pt]{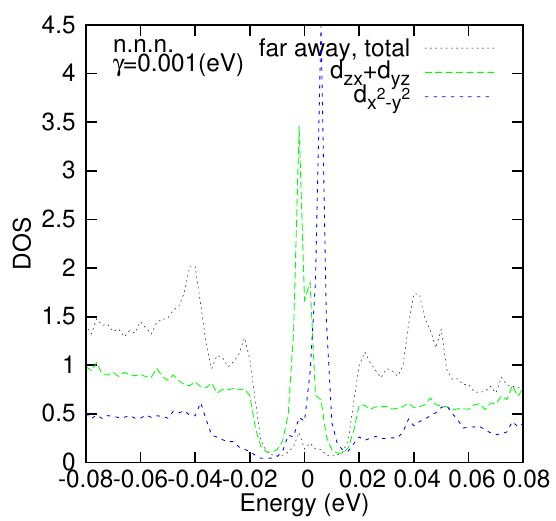}}
  \subfigure[3rd n.n.]{\includegraphics[width=120pt]{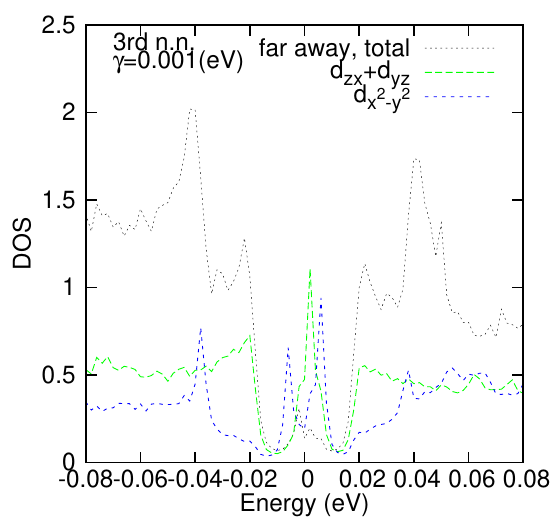}}
  \caption{Partial LDOS in the case of $I=1.0$ eV on the several sites
  near the impurity.  (a) shows
  partial LDOS at just on the impurity site, (b) on the nearest, (c) the
  next-nearest, and (d) the third-nearest-neighbor site, respectively. Lines in
  each figure represent $\rho_{1}(\bm{r},\omega)+\rho_{2}(\bm{r},\omega)$
  (sum of the contributions from $d_{zx}$ and $d_{yz}$ orbitals) and
  $\rho_{3}(\bm{r},\omega)$ (contribution from $d_{x^2-y^2}$ orbital). LDOS at
  a site far from the impurity is also plotted.}
  \label{fig2}
 \end{center}
\end{figure}

\begin{figure*}[t]
 \begin{center}
  \subfigure[$I=0.5$ eV]
  {\includegraphics[width=110pt]{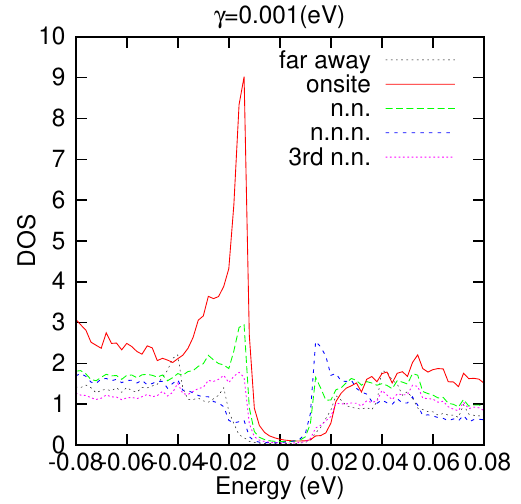}}
  \subfigure[$I=1.0$ eV]
  {\includegraphics[width=110pt]{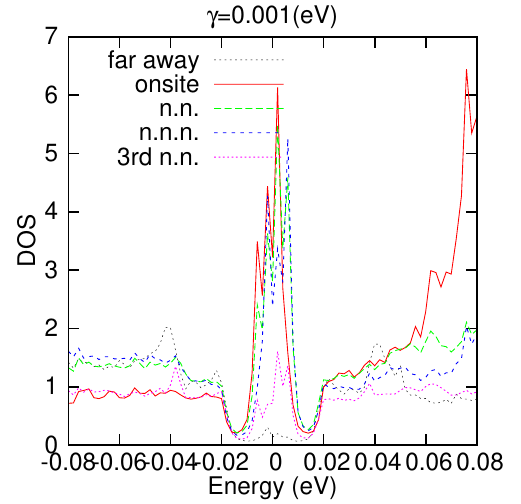}}
  \subfigure[$I=2.0$ eV]
  {\includegraphics[width=110pt]{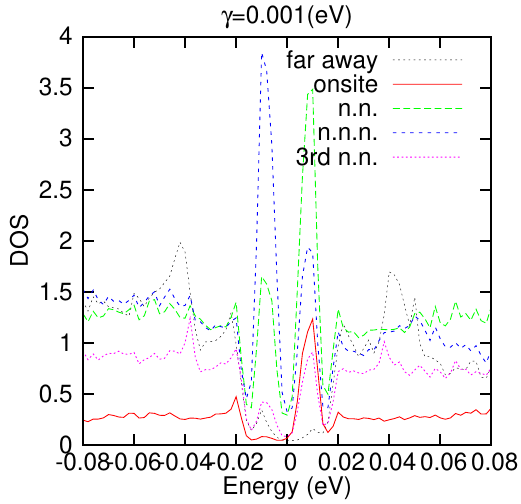}}
  \subfigure[$I=8.0$ eV]
  {\includegraphics[width=110pt]{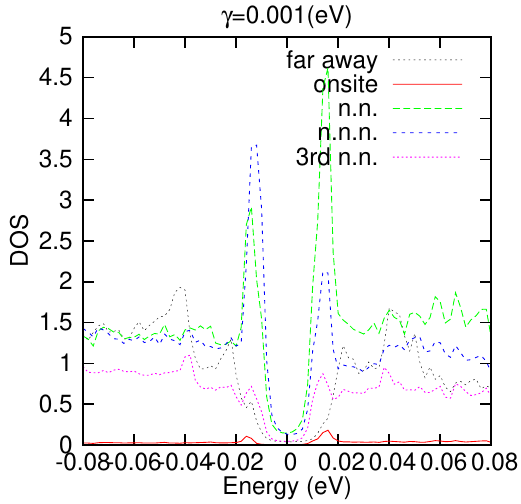}}\\
  \subfigure[$I=-0.5$ eV]
  {\includegraphics[width=110pt]{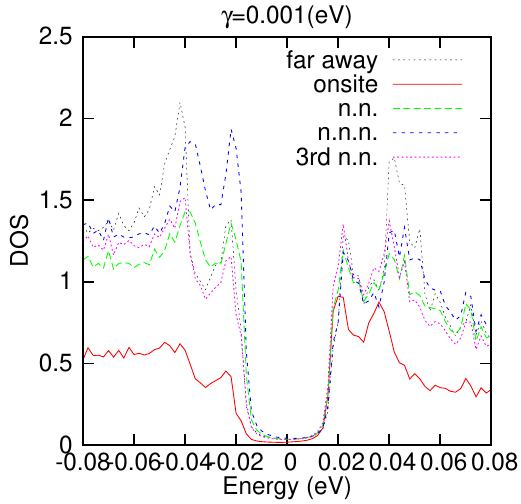}}
  \subfigure[$I=-1.0$ eV]
  {\includegraphics[width=110pt]{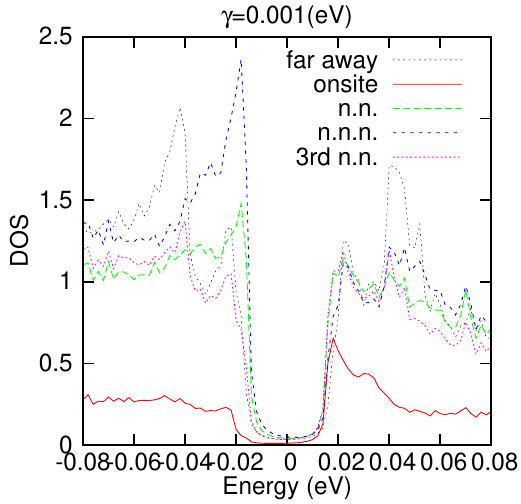}}
  \subfigure[$I=-2.0$ eV]
  {\includegraphics[width=110pt]{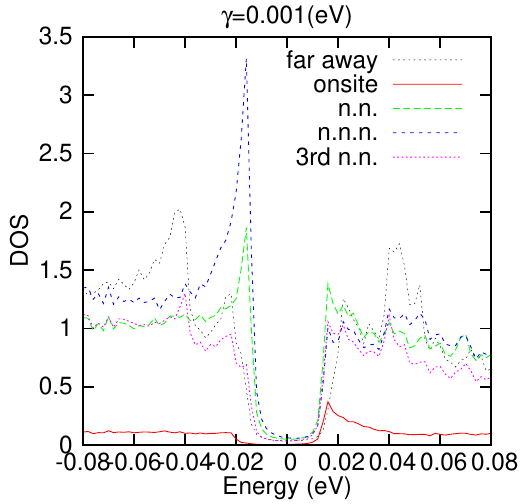}}
  \subfigure[$I=-8.0$ eV]
  {\includegraphics[width=110pt]{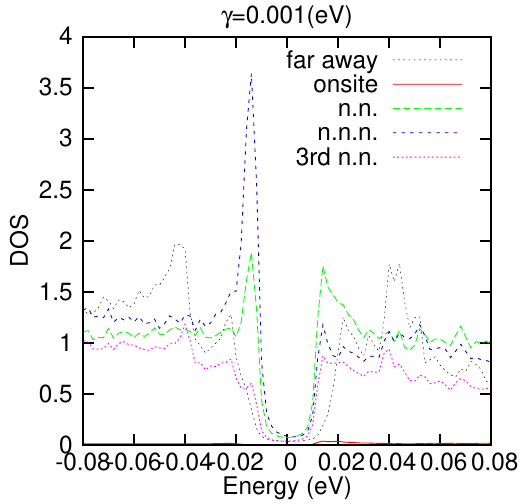}}
  \caption{LDOS for various impurity potential $I$. Lines in each figure
  represents LDOS at the site far from the impurity site, on the
  impurity site, nearest neighbor site, next nearest neighbor site, and
  3rd nearest neighbor site.}
  \label{fig3}
 \end{center}
\end{figure*}

Figure \ref{fig1} shows the obtained order parameter and its modulation
around the
impurity site. Here, we consider the case of $I=1.0$ eV. A representative
order parameter, $\Delta^{\mathrm{n.n.n.}}_{aa}(\bm{r})$, defined as
\begin{equation}
 \Delta^{\mathrm{n.n.n.}}_{aa}(\bm{r})
  \equiv\sum_{\bm{r}'}{}'\Delta_{\bm{r}'a,\bm{r}a}
  =\sum_{\bm{r}'}{}'\sum_{a'}g^{(2)}_{aa'a'a}
  \langle
  c^\dagger_{\bm{r}'a'\sigma}c^\dagger_{\bm{r}a'\bar{\sigma}}
  \rangle,
  \label{Deltannn}
\end{equation}
is plotted in Fig.~\ref{fig1}. Here, $\sum'_{\bm{r}'}$ means that the
summation over the next nearest neighbor sites of $\bm{r}$. 
The amplitude of $\Delta^{\mathrm{n.n.n.}}_{aa}(\bm{r})$ is suppressed
near the impurity and we can estimate the coherence length to be about
four or five lattice spacing. 
At a site far away from
the impurity, where the bulk behaviors are expected, we have 
$\Delta^{\mathrm{n.n.}}_{11}/\Delta^{\mathrm{n.n.}}_{33}\sim -0.41$,
$\Delta^{\mathrm{n.n.n.}}_{11}/\Delta^{\mathrm{n.n.}}_{33}\sim -0.30$,
$\Delta^{\mathrm{onsite}}_{33}/4\Delta^{\mathrm{n.n.}}_{33}\sim 0.54$, and
$\Delta^{\mathrm{n.n.n,}}_{33}/\Delta^{\mathrm{n.n.}}_{33}\sim -0.50$. 
Roughly speaking, these ratios are consistent with those obtained in
RPA for $n=6.1$ and $J_{H}/U=0.2$\cite{JPSJ.79.033703}. Although the
value of $\Delta^{\mathrm{n.n.n.}}_{11}/\Delta^{\mathrm{n.n.}}_{33}$ is
slightly overestimated compared with RPA, the present result captures
most of the features of RPA result.
$\Delta^{\mathrm{n.n.}}_{aa}(\bm{r})$, which is defined for the
nearest-neighbor sites, shows a similar behavior as
$\Delta^{\mathrm{n.n.n.}}_{aa}(\bm{r})$, while
$\Delta^{\mathrm{onsite}}_{aa}(\bm{r})\equiv \Delta_{\bm{r}a,\bm{r}a}$
shows an overshooting behavior, i.e., it exceeds the bulk value at some
points around the impurity.

Next, we show LDOS around the impurity for the case of $I=1.0$ eV
in Fig.~\ref{fig2}. In each figure, the partial LDOS
for $d_{zx/yz}$ and $d_{x^2-y^2}$ orbitals are plotted. From 
Figs.~\ref{fig2}(a)-\ref{fig2}(c), we can see the clear formation of the
impurity induced in-gap bound state around impurity. 
When we look at the obtained spectrum more closely, 
we find that the peaks appear as a pair, i.e., at $\pm E$ for each
orbital, and that the width of each pair, $E$, depends on the
orbital. As a result, total LDOS, which is the sum of the partial LDOS,
shows novel multiple peak structures (see also Fig.~\ref{fig3}(b)). This
feature is characteristic in iron pnictides and can be captured only
in our realistic five-orbital model.
At the third nearest neighbor site (Fig.~\ref{fig2}(d)), on the other
hand, the in-gap bound 
state peaks become small and the bulk-like coherence peaks recover. 

Impurity potential dependence of the LDOS is summarized in
Fig.~\ref{fig3} where the results for $I$ ranging from
$-8.0$ eV to $+8.0$ eV are shown. In each figure, we plot LDOS at
several sites around the impurity up to third nearest neighbor site as
well as LDOS far away from the impurity. As a function of $I$ ($I>0$),
(Fig.~\ref{fig3}(a-d)), we find that the bound state appears at the edge
of the gap ($I=0.5$ eV), and moves toward nearly zero energy ($I=1.0$ eV), and
then, goes back to the gap edge ($I=2.0$ eV and $I=8.0$ eV). As a
result, the bound state formation is most prominent at $I=1.0$ eV. On
the other hand, when $I$ is negative, (Fig.~\ref{fig3}(e-h)), the
bound-state energies stick to the gap edge and do not approach the zero
energy. In this case, the amplitude of the peak just grows with
increasing $|I|$. 
Comparing these cases, we can see that the impurity with negative $I$ has
much less effects than the impurity with positive $I$ when $|I|$ is
relatively small. In contrast, when $|I|$ is large and in the unitary
limit, LDOS structures are similar for both positive and negative
$I$'s. This is reasonable since the impurity with large $|I|$ works as a
site onto which electrons cannot hop, whichever the sign of $I$ is. 

Experimentally, LDOS can be measured by scanning tunneling
microscopy/spectroscopy (STM/STS). Since the in-gap bound state around
a non-magnetic impurity does not
appear in s$_{++}$ state, the bound state formation discussed above can
be used to distinguish s$_{+-}$ and s$_{++}$ state. In particular, if
the impurity potential is around $I\sim 1.0$ eV, the difference between
s$_{+-}$ and s$_{++}$ states becomes most prominent since
the in-gap bound state appears at the near zero-energy and the peak is
very large. Even if the large in-gap state is not observed, the spectrum
at the gap edge can be carefully investigated to distinguish the
s$_{+-}$ and s$_{++}$ state. Up to now, the superconducting gap has been
successfully observed in Fe(TeSe)
system\cite{PhysRevB.80.180507,hanaguri}, and the
detailed comparison of the theory and experiments is an interesting
future work. 

\begin{figure}[tb]
 \begin{center}
  \subfigure[$I=1.0$, $\omega=-0.002$ eV]
  {\includegraphics[width=120pt]{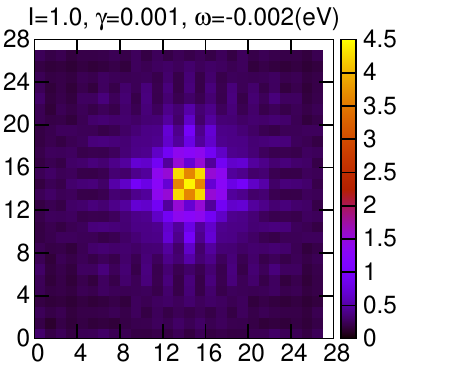}}
  \subfigure[$I=8.0$, $\omega=-0.014$ eV]
  {\includegraphics[width=120pt]{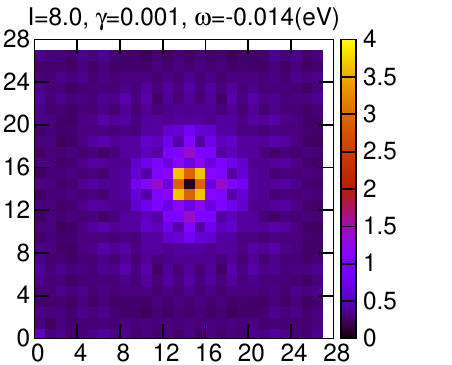}}
  \caption{Real space map of LDOS for (a) $I=1.0$, $\omega=-0.002$ eV
  and (b) $I=8.0$, $\omega=-0.014$ eV.}
  \label{fig4}
 \end{center}
\end{figure}

We plot the real space map of LDOS in Fig.~\ref{fig4}. In order
to see the typical behavior of the impurity induced bound states, we
choose the two cases with (a) $I=1.0$ eV, $\omega=-0.002$ eV
and (b) $I=8.0$ eV, $\omega=-0.014$ eV, in which the bound state
peaks in LDOS are clearly seen. For both cases, we find that the real
space distribution of LDOS or the extension of the bound state is nearly
isotropic. This will be due to the following two reasons: 1) Fermi
surfaces of the used model are circular, and 2) the obtained gap
function is nodeless in the bulk limit, which means that there is no
special direction associated with the node. This feature will be changed
if we consider a parameter region where the nodal gap function is
obtained, or if the square Fermi surfaces appear. 

Here, we discuss the obtained results in connection to the 
$T$-matrix approximation, which has been used to study the formation of
the bound state in the superconducting
state\cite{shiba,RevModPhys.78.373}. 
This approximation applied on the similar five orbital model
gives the results consistent with the present paper\cite{Onari:2009},
although the site-resolved information is difficult to obtain in the
$T$-matrix approximation. 
We discuss the following two points.  
One is about the $I$ dependence of the spectrum, which is understood
from the particle-hole asymmetry of DOS in the present model. To be more
specific, the DOS has larger weight below the Fermi energy than above
the Fermi energy. We can show that this asymmetry results in the
formation of the nearly zero-energy bound state when $I$ is small and
positive. (Details will be published elsewhere.) For other values of
$I$, we can show similarly that the bound states are formed at the gap edge. 
The second point is about the multiple peak structure found in 
the case of $I=1.0$ eV. In this case, we can show that the $T$-matrix
can be block diagonalized with each block having only intra-orbital
contribution. This means that each orbital can sustain the bound state
independently, and this is the origin of the multiple peak structure.

It is tempting to speculate that the present results are connected to
the experimentally observed robustness against the impurity
doping. Namely, the impurity effect is weak when $I$ is negative and
small, which may explain the robustness against the impurity doping,
although the magnitude of $I$ and the relation to the residual
resistivity should be taken account of carefully\cite{Onari:2009}. 

In summary, we have calculated the LDOS around a non-magnetic impurity
in the effective model for iron-pnictide superconductors. The model used
here has realistic band structure and the effective interaction is
determined so as to reproduce the RPA results of the gap function. It
has been shown that the observation of the in-gap bound state enables us
to distinguish the s$_{++}$ and s$_{+-}$ states. An impurity with $I\sim
1.0$ eV gives large low energy peak, and we have also shown that the
bound state peak appears at just below (or above) the gap edge in the
unitary limit. An impurity with negative and small $I$ has relatively
small effects on LDOS. The modeling of the case having the nodal gap
function and of the recently proposed phonon-assisted
orbital-fluctuation-medeated superconductivity are interesting
extensions of the present study. 
Similar analysis on the quasi-particle interference patterns would also
give important information.

\section*{Acknowledgment}
We thank K.~Kuroki, H.~Ikeda, K.~Nakamura, and T.~Hanaguri for
stimulating discussion. T.K. is supported by JSPS Research Fellowship.

\end{document}